\documentclass[letter]{aa}

\usepackage{amsmath}
\usepackage{graphicx}
\usepackage{changepage}
\usepackage{listings}
\usepackage{siunitx}
\usepackage{natbib}
\usepackage{hyperref}
\usepackage{multirow}
\usepackage{float}
\usepackage{placeins}
\usepackage{pdflscape}  
\usepackage{longtable}  
\usepackage{lastpage}
\usepackage{txfonts}

\usepackage{xcolor}

\begin{document}

   \title{Surface Nitrogen enrichment from the LMC to the Early Universe: the roles of rotation, overshooting and mass loss}
   \titlerunning{Surface Nitrogen enrichment from the LMC to the Early Universe}

\author{{Gautham N. Sabhahit\inst{\ref{AOP}}}
    \and 
       {Jorick S. Vink\inst{\ref{AOP}}}
    }

\institute{
   {Armagh Observatory and Planetarium, College Hill, Armagh BT61 9DG, N. Ireland\label{AOP}}
   \email{gauthamns96@gmail.com}
             }


 
\abstract
{The strong nitrogen (N) features observed in high-redshift galaxies, such as GN-z11, with JWST point to an efficient source of N production in the early Universe. Very massive stars (VMSs) are promising candidates owing to their strong winds and large convective cores of CNO-processed material. Whether massive-star surfaces in general become N-enriched depends on the interplay of convective-boundary mixing (overshooting), rotation, and mass loss, which remains poorly constrained.
Using a grid of MESA evolutionary models, we isolate the overshooting and rotation channels at $40$, $100$, and $200\,M_\odot$ and investigate how these channels bring N-rich material to the surface of massive and very massive stars as a function of initial mass and metallicity. We also construct a synthetic Large Magellanic Cloud (LMC) population based on a larger initial mass grid ranging from $20-300\,M_\odot$ for comparison with the observed N-enriched fraction.
VMS surface enrichment is remarkably robust across cosmic metallicity. At lower metallicities, wind stripping is weaker and the overshooting channel shuts off entirely, yet a modest amount of rotation of $\sim90\,\mathrm{km\,s^{-1}}$ alone still enriches the surface. At higher metallicity, stronger winds dominate and uncover the core almost independently of the adopted mixing. The two channels operate through physically distinct mechanisms: overshooting enriches the surface only once winds strip the envelope down into the N-rich layers left by the receding core, giving an abrupt jump. Rotation, by contrast, mixes N outward continuously through the layers above the core, producing a gradual rise. Applied to the LMC, a population with modest rotation drawn from the observed velocity distribution and modest overshooting reproduces the observed N-enriched fraction without large core boundary mixing efficiencies otherwise required by non-rotating models.
}


   \maketitle
%

\section{Introduction}
\label{sec: intro}

Massive stars are efficient factories of nitrogen (N): hydrogen burning through the CNO cycle converts most of the initial carbon (C) and oxygen (O) in the convective core into $^{14}$N. Any process that connects this processed core to the surface -- envelope stripping or enhanced mixing -- leaves a distinctive imprint of N enhancement and C depletion~\citep{Langer2012}, and once exposed, this material can be swept out by winds to enrich the surrounding interstellar medium (ISM).

Very massive stars (VMSs, $M_\mathrm{init} > 100\,M_\odot$) are especially promising in this respect. Born close to the Eddington limit, they launch strong winds~\citep{GH2008, Vink2011, VinkR} capable of exposing their large CNO-processed convective cores. 
For example, the Sunburst Arc~\citep{Mestric2023} requires several hundred $M_\odot$ of N to explain its observed enrichment, and \citet{Vink23} showed that the N ejected by a few hundred VMSs inferred in this galaxy would agree with this, as VMS eject of order a solar mass of N into the ISM \citep{Sabhahit2022,Higgins2023}. Interest in this channel has been renewed by the James Webb spectroscopy of the early Universe, which has uncovered high-redshift galaxies with unexpectedly strong N features, most notably GN-z11 at $z=10.6$~\citep{Bunker2023, Senchyna2024}. Crucially, even at the lower metallicities ($Z$) where winds weaken, VMSs can still enrich their surroundings~\citep{Vink23, Higgins2025}.

In the Local Universe we have the added advantage of resolving individual stars and analysing them with detailed wind-atmosphere codes, making their surface abundances a primary observational constraint on internal mixing. Such analyses now span hundreds of O and WNh stars in the nearby Large Magellanic Cloud (LMC), through the VLT-FLAMES Tarantula Survey~\citep{Evans2011, Best2014} and UV HST/STIS spectroscopy of the R136 core~\citep{Crowther2016, Best2020, Brands22}, and reveal that a substantial fraction are already surface N-enriched on the main sequence.

Reproducing this is not straightforward, as several processes transport N outward: rotationally-induced instabilities mix core material through the radiative envelope~\citep{Heger2000}, while convective-boundary mixing enlarges the core and exposes processed layers once the envelope is stripped. A significant fraction of massive stars also undergo binary interaction~\citep{Sana2012, Sana2013}, where mass transfer strips the donor and accretion or mergers spin up and mix N outward~\citep{Langer2012, deMink2013}. These channels are partly degenerate, so the same enrichment can be matched by very different combinations, and disentangling them requires models that isolate each mechanism.

\begin{figure*}    
\includegraphics[width = 0.96\textwidth]{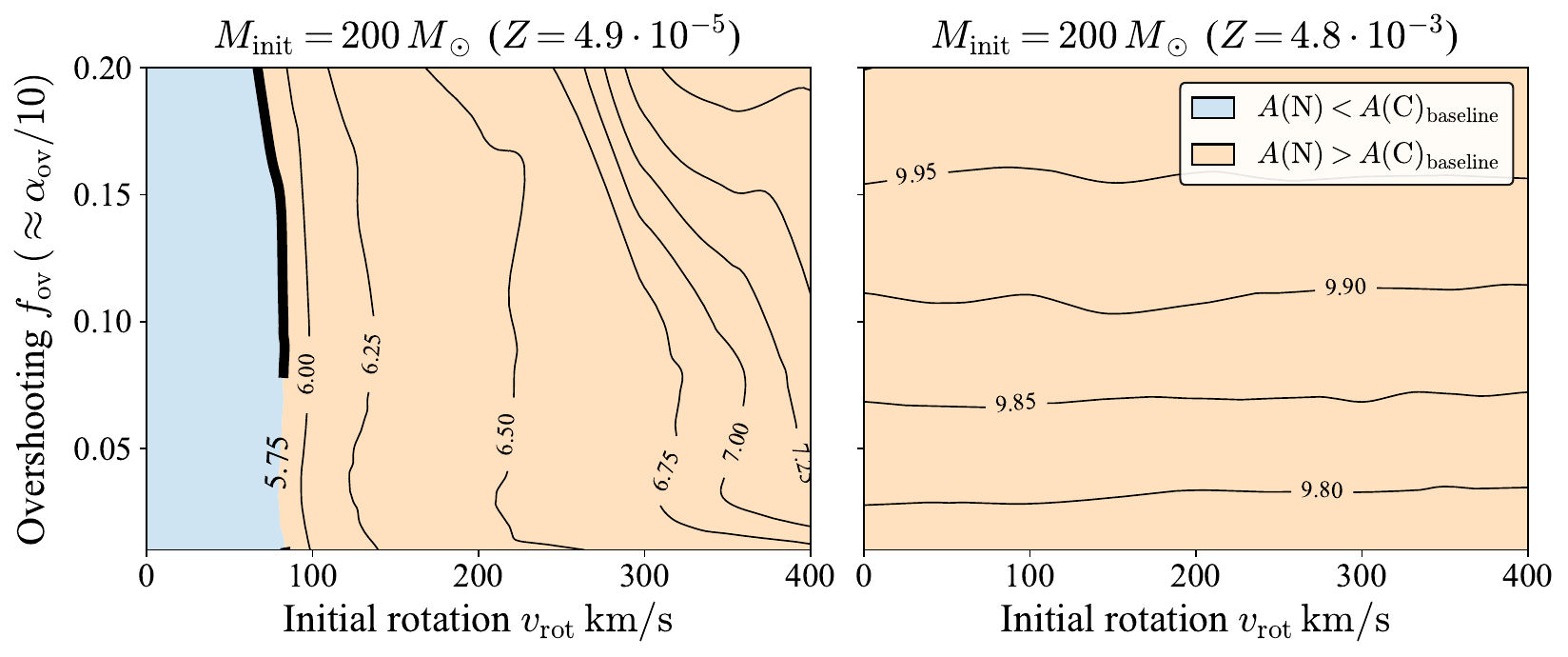}    
\caption{Maximum surface nitrogen abundance $A(\mathrm{N})$ reached by a $200\,M_\odot$ star as a function of overshooting and rotation, for two metallicities: $Z=4.9\cdot10^{-5}$ (\textit{left}) and $Z=4.8\cdot10^{-3}$ (\textit{right}). Only models hotter than $T_\mathrm{eff} = 30\,\mathrm{kK}$, corresponding to hot O stars, are considered when finding the maximum. Shading is blue where $A(\mathrm{N})$ stays below the baseline C abundance $A(\mathrm{C})_\mathrm{baseline}$ and orange where it exceeds it, with $A(\mathrm{C})_\mathrm{baseline} = 5.75$ (left) and $7.75$ (right). Abundances are given as $A(\mathrm{el}) = 12+\log(n_\mathrm{el}/n_\mathrm{H})$.}  
\label{fig: 200}
\end{figure*}

Recently, \citet{Marchant2026} modelled the surface N-enrichment probability in the Tarantula Nebula of the LMC with a population synthesis of non-rotating models, and found that reproducing the observations below $\log(L/L_\odot)\approx6.5$ required strong core-boundary mixing, with overshooting efficiencies $\alpha_\mathrm{ov}=1$--$3$.

In this Letter we revisit the problem with rotation included. We first show that the overshooting channel relies entirely on the stripping of the envelope layers to transport N from the core to the surface, and is therefore not viable at very low $Z$ where the winds are too weak. Enriching the surface instead requires some form of mixing in the radiative layers beyond the overshooting region which in our case is provided by rotation-induced mixing. We then show that VMS surface N enrichment is robust against this, since it proceeds through wind stripping at high metallicity and survives through modest rotational mixing even at very low $Z$ in the absence of mass loss (Sect.~\ref{sec: VMS}). Finally, we show that modest rotation drawn from the observed velocity distribution reproduces the LMC N enrichment probabilities without strong enlargement of the convective core (Sect.~\ref{sec: popsynth}).

\section{The role of very massive stars in nitrogen enrichment}
\label{sec: VMS}

\subsection{Stellar evolution models}
\label{sec: models}
We computed a grid of massive and VMS models with the MESA stellar evolution code~\citep{MESA11}, covering the initial mass range $20$--$300\,M_\odot$. Stars are evolved from the pre-MS to the end of the MS ($X_\mathrm{c} < 0.01$) for two different metallicities: $Z=4.8\cdot10^{-3}$ following LMC-like abundances (see Appendix~\ref{app: mesa_details}), and a factor of 100 lower with $Z=4.9\cdot10^{-5}$.\footnote{This is roughly $Z_\odot/300$ where $Z_\odot = 0.013$ from \citet{Asplund2005}.}

Convective boundary mixing across the convection–radiation boundary is treated with overshooting, in which convective blobs retain a non-zero velocity at the formal Schwarzschild boundary and can therefore overshoot into the radiative region. We use an exponentially decaying overshooting formalism following~\citet{Herwig2000}, controlled by the free parameter $f_\mathrm{ov}$ which we vary from 0.01 to 0.2. The parameter $f_\mathrm{ov}$ relates to the step-overshooting parameter by $\alpha_\mathrm{ov} \approx 10\,f_\mathrm{ov}$~\citep{Moravveji2015}, so that e.g. $f_\mathrm{ov}=0.03$ and $\alpha_\mathrm{ov}=0.3$ reproduce similar MS widths and lifetimes. We report $f_\mathrm{ov}$ throughout, from which the approximate $\alpha_\mathrm{ov}$ follows by multiplying by a factor of ten.

Rotational mixing is included through the rotationally-induced instabilities of~\citet{Heger2000}, which transport both material and angular momentum from core to surface~\citep{Maeder2009}. Initial rotational velocities range from 0 to 400 $\mathrm{km\,s^{-1}}$. We additionally test the impact of magnetic core--envelope coupling by switching the Spruit--Taylor ~(ST) dynamo\citep{Spruit1999, Spruit2002} on and off.

For mass loss we use the rates of~\citet{Sabhahit2022}, which switch from optically-thin O-type rates~\citep{Vink2001} to an optically-thick WR-type scaling based on~\citet{Vink2011}. Since recent empirical and theoretical work suggests a downward revision of the~\citet{Vink2001} O-star rates in the $\sim 30\,M_\odot$ range~\citep{Bjorklund2021, Krticka21, Hawcroft2021}, we also test the effect of reducing them by a factor of three.

Throughout this work, a model is classified as surface N-rich, using the same definition as~\citet{Marchant2026}, when the surface N abundance rises beyond the baseline C abundance.

\subsection{Very massive stars at high and low metallicity}
\label{sec: VMS_Z}

VMSs are especially promising candidates for N enrichment because of their large convective-core-to-total-mass ratio. For initial masses above $\sim200\,M_\odot$, more than 90\% of the stellar mass resides in the convective core~\citep{Yusof2013, Sabhahit2022}, where N produced by the CNO cycle is almost instantly fully mixed. In principle, winds only need to strip $\sim10\%$ of the envelope to expose this material and enrich the ISM~\citep{Vink23}.

\begin{figure*}    
\includegraphics[width = 0.96\textwidth]{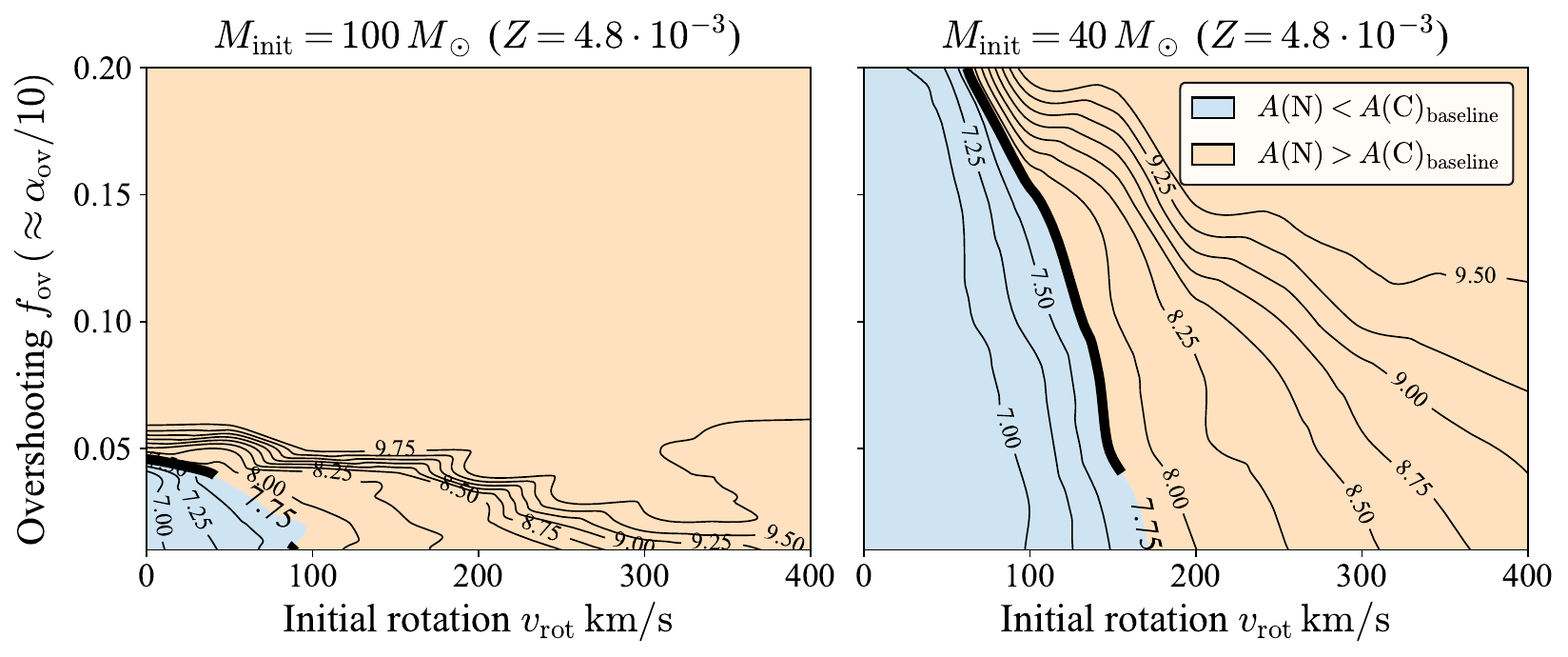}    
\caption{Maximum surface nitrogen abundance $A(\mathrm{N})$ of a $100\,M_\odot$ \textit{(left)} and a $40\,M_\odot$ \textit{(right)} star as a function of overshooting and rotation, at fixed metallicity $Z=4.8\cdot10^{-3}$. Symbols and shading are as in Fig.~\ref{fig: 200}.} 
\label{fig: 40_100}
\end{figure*}

However, in reality, a large convective core on its own is not sufficient, because the N-rich layers must actually be uncovered. In Fig.~\ref{fig: 200} (left sub-panel) we show the maximum surface N abundance reached by a $200\,M_\odot$ model computed with zero mass loss appropriate for very low--$Z$. We find that with neither wind stripping nor rotation, the models fail to become N-enhanced regardless of the overshooting adopted. Overshooting acting alone is therefore \emph{not} a viable channel and is of no consequence if the winds never strip down into the core layers. 

Instead, at such low $Z$, bringing N to the surface requires mixing in the radiative layers beyond the overshooting region. In our models this mixing is provided by rotation, and a modest $\sim90\,\mathrm{km\,s^{-1}}$ already suffices to bring N to the surface. This provides the necessary surface enrichment, although subsequent ejection is still required for the material to reach the surrounding ISM. Here rotation may help again, since VMSs sit close to the $\Omega\Gamma$ limit~\citep{MM2000, Winch2025}, the Eddington limit modified by rotation, and may therefore remain efficient N producers in the $Z$-poor early Universe. 

At higher metallicities ($Z \gtrsim 0.1\,Z_\odot$), wind stripping dominates. Fig.~\ref{fig: 200} (right sub-panel) shows the surface N enrichment as a function of $f_\mathrm{ov}$ and $\varv_\mathrm{rot}$ computed with the~\citet{Sabhahit2022} rates. The winds completely dominate, such that the surface is enriched across the \emph{entire} parameter space regardless of the adopted mixing.

\subsection{The overshooting and rotation channels at 40 and 100\,$M_\odot$}
\label{sec: 40_100}

Overshooting and rotation enrich the surface through fundamentally different mechanisms, but the two are virtually impossible to decouple when winds dominate (e.g. the $200\,M_\odot$ case above). We therefore isolate them at lower masses in Fig.~\ref{fig: 40_100}.

The overshooting channel has two key properties. First, enrichment arises because the convective core (including the overshoot layers) recedes during the MS, leaving behind fossil layers of CNO-enriched material above it. Mass loss must then strip the envelope down to these layers before the N is exposed, so this channel depends entirely on wind strengths (Sect.~\ref{sec: VMS_Z}). Second, because the core composition is nearly uniform, the surface rise from the baseline to the CNO-processed value is very abrupt.

Rotation, by contrast, drives instabilities in the radiative layers \emph{above} the core and overshooting region, directly connecting the core to the surface, so the enrichment it produces is gradual rather than step-like. This gradual build-up was shown by~\citet{Higgins2019} to be necessary to reproduce the 3:1 N abundance ratio between the components of the eclipsing binary HD\,166734, which no non-rotating, overshooting-only model could match.

Comparing the $\varv_\mathrm{rot} = 0$ models at varying overshooting in Fig.~\ref{fig: 40_100}, we find that the $40\,M_\odot$ model is not enriched even at $f_\mathrm{ov} = 0.2$, whereas $f_\mathrm{ov} = 0.05$ already suffices at $100\,M_\odot$. The same trend holds for rotation: N enrichment occurs for $\varv_\mathrm{rot} \gtrsim 150\,\mathrm{km\,s^{-1}}$ at $40\,M_\odot$ but already for $\gtrsim 75\,\mathrm{km\,s^{-1}}$ at $100\,M_\odot$. The larger the relative core size, the less mixing is needed, which is precisely what makes the surfaces of VMSs so easily enrichable.

\section{Population synthesis and comparison with observations}
\label{sec: popsynth}

We construct a synthetic population by sampling initial masses from a Salpeter Initial Mass Function \citep[IMF, $n=-2.35$;][]{Salpeter55}, and ages with a constant star formation rate (SFR) over the last $10\,\mathrm{Myr}$. A more realistic star formation history for the LMC~\citep{Schneider2018b} has minimal impact on the N-enriched fraction~\citep{Marchant2026}, so we adopt a constant SFR. Since births are sampled uniformly, a model whose lifetime is shorter than its assigned age will already have died by the present and is discarded. Initial rotational velocities are sampled over $0-400\,\mathrm{km\,s^{-1}}$ following~\citet{RA2013}, whose distribution peaks at $\varv_\mathrm{e} = 100\,\mathrm{km\,s^{-1}}$, and we fix $f_\mathrm{ov} = 0.03$. For a fair comparison with the observed O and WNh stars, we reject any model once it evolves below $T_\mathrm{eff} = 30\,\mathrm{kK}$.

\begin{figure}    
\includegraphics[width = \columnwidth]{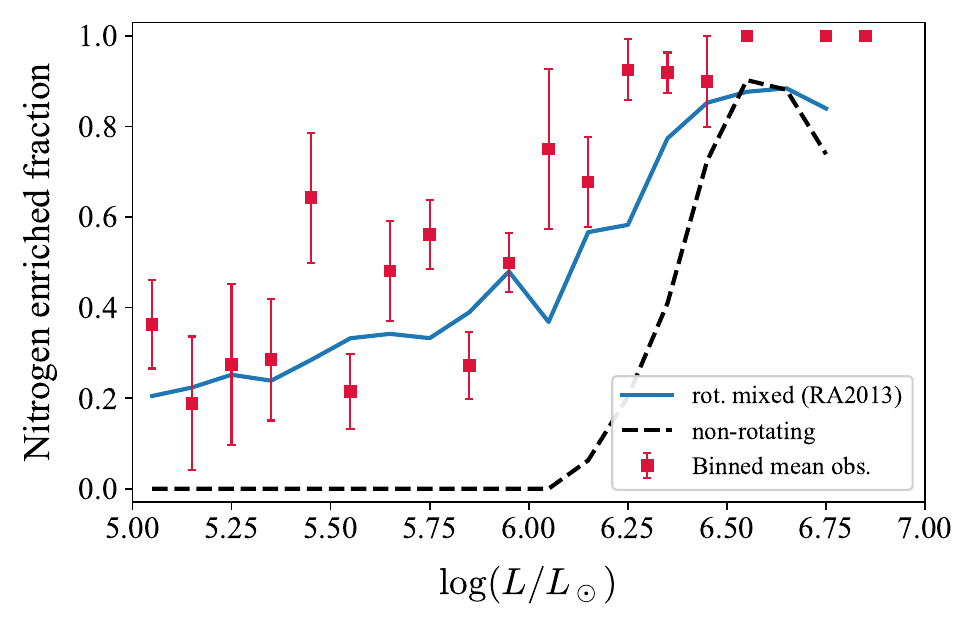}    
\caption{Nitrogen enrichment probability as a function of luminosity. Red squares mark the mean observed probability, binned in $\Delta\log(L/L_\odot) = 0.1$\,dex, from the LMC sample compiled by~\citet{Marchant2026}, compared against population synthesis results with no rotation (dashed black) and with rotation sampled from the~\citet{RA2013} distribution, which peaks at $100\,\mathrm{km\,s^{-1}}$ (solid blue).}  
\label{fig: nitrogen_enriched_fraction}
\end{figure}

Figure~\ref{fig: nitrogen_enriched_fraction} shows the resulting N-enriched fraction as a function of luminosity, compared against the LMC probabilities of~\citet{Marchant2026}, who use a Bayesian model built on the O and WNh stars in the VLT-FLAMES Tarantula Survey~\citep{Evans2011, Best2014} and the R136 stars~\citep{Crowther2016, Best2020}.

The non-rotating population clearly fails to reproduce the observed probabilities, as already noted by~\citet{Marchant2026}, most evidently over $\log(L/L_\odot) = 5$--$6$ where the models show essentially no surface enrichment. This follows directly from overshooting alone at $f_\mathrm{ov} = 0.03$ being unable to enrich the surface (Sect.~\ref{sec: 40_100}).

\citet{Marchant2026} proposed that this disagreement below $\log(L/L_\odot) = 6.5$ could be reconciled by enlarging the core-boundary mixing to $\alpha_\mathrm{ov} = 1$--$3$. The essential ingredient, however, is some form of mixing in the radiative layers above the overshoot region (Sect.~\ref{sec: 40_100}). We employ rotational mixing, although other mechanisms such as gravity waves could serve the same role. Modest rotation from the~\citet{RA2013} distribution, combined with modest overshooting ($f_\mathrm{ov} = 0.03$), reproduces the observations with an overshoot efficiency an order of magnitude below that required by~\citet{Marchant2026}.

We perform several additional tests in Appendix~\ref{app: rot_velo} to verify that the evolved surface velocities remain consistent with the~\citet{RA2013} input, i.e. the models do not spin up to unobserved velocities, under reduced O-star rates and with the ST dynamo active.

\section{Concluding remarks}
\label{sec: conclusions}
In this work we examined the interplay between overshooting, rotation and mass loss in the surface N enrichment of massive and very massive stars, and confronted a synthetic LMC population with the observed surface N-enriched fraction. Our main findings are:

\begin{itemize}

\item VMSs ($\gtrsim100\,M_\odot$) enrich their surfaces almost inevitably. With large convective-core-to-total mass ratios and strong winds, they need only shed a few tens of $M_\odot$ to uncover CNO-processed material, which holds for $Z \gtrsim Z_\odot/10$ and is therefore relevant both for Local Universe clusters such as the Tarantula Nebula and for early Universe galaxies such as GN-z11. Even at very low $Z$ ($\lesssim Z_\odot/100$), where mass loss may be negligible, the surface enrichment persists, as a modest $\sim90\,\mathrm{km\,s^{-1}}$ rotation suffices to bring N to the surface.

\item Overshooting and rotation enrich through different mechanisms. For overshooting alone to enrich the surface, the envelope must be sufficiently stripped; this channel shuts off entirely when the winds are too weak to dig into the N-enriched fossil layers of the core.

\item Reproducing the observed LMC N-enriched probabilities requires mixing in the layers directly above the overshooting region. In our models rotationally-induced mixing supplies this, driving continuous outward transport and a gradual rise in surface N -- the same gradual enrichment required by~\citet{Higgins2019} to match the components of the eclipsing binary HD\,166734.

\end{itemize}

While we have benchmarked our models against the LMC, where individual stars can be resolved, high-redshift N emitters such as GN-z11~\citep{Bunker2023, Senchyna2024} can only be studied through their integrated light. Locally, surveys of massive stars in more metal-poor environments, such as the SMC~\citep{Shenar2024} and dwarf galaxies of even lower $Z$, would test whether rotational mixing still enriches the surface where winds are weak. At high redshift, where VMS winds remain uncertain and need to be properly modelled, including VMSs in population synthesis and chemical evolution calculations~\citep{Vink23} would allow their predicted N/O ratios to be compared with JWST spectra~\citep{Cameron2023}. Beyond N, the same mixing that exposes CNO-processed material at low $Z$ would also bring up products of hotter hydrogen burning, such as Na~\citep{BastianLardo2018}, linking N emitters to the abundance anomalies of globular clusters. And since weaker winds at low $Z$ may allow stars to form at even higher masses~\citep{Vink2018}, establishing whether rotational mixing remains efficient in this regime is essential for understanding early chemical enrichment.

\begin{acknowledgements}
GNS and JSV are supported by STFC funding under grant number ST/Y001338/1. 
\end{acknowledgements}

\bibliographystyle{aa}
\bibliography{References.bib}

\appendix

\section{Rotational velocity distribution and additional tests}
\label{app: rot_velo}

In Sect.~\ref{sec: popsynth} we presented population synthesis models built on MESA evolutionary tracks, with Salpeter IMF weighting, a constant SFR history, and an initial rotation distribution following~\citet{RA2013}. One of the main objections raised by~\citet{Marchant2026} to the use of rotating models is that the large majority of O stars in the VFTS sample are slow rotators, with the single-star rotation distribution peaking at $\sim100\,\mathrm{km\,s^{-1}}$ and only a sparsely-populated high-velocity tail~\citep{RA2013}.

While our initial distribution is, by construction, dominated by objects near $100\,\mathrm{km\,s^{-1}}$, we note that~\citet{RA2013} is a present-day distribution rather than an initial one, and converting between the two is inherently model dependent. We therefore take it as our initial distribution and check a posteriori that the evolved velocities do not drift away from it, since the stars could in principle spin up and over-populate the high-velocity tail that is not observed. We also run additional models to test the sensitivity of our results to the adopted mass-loss rates and to a magnetic coupling between the core and the envelope.

\begin{figure}    
\includegraphics[width = \columnwidth]{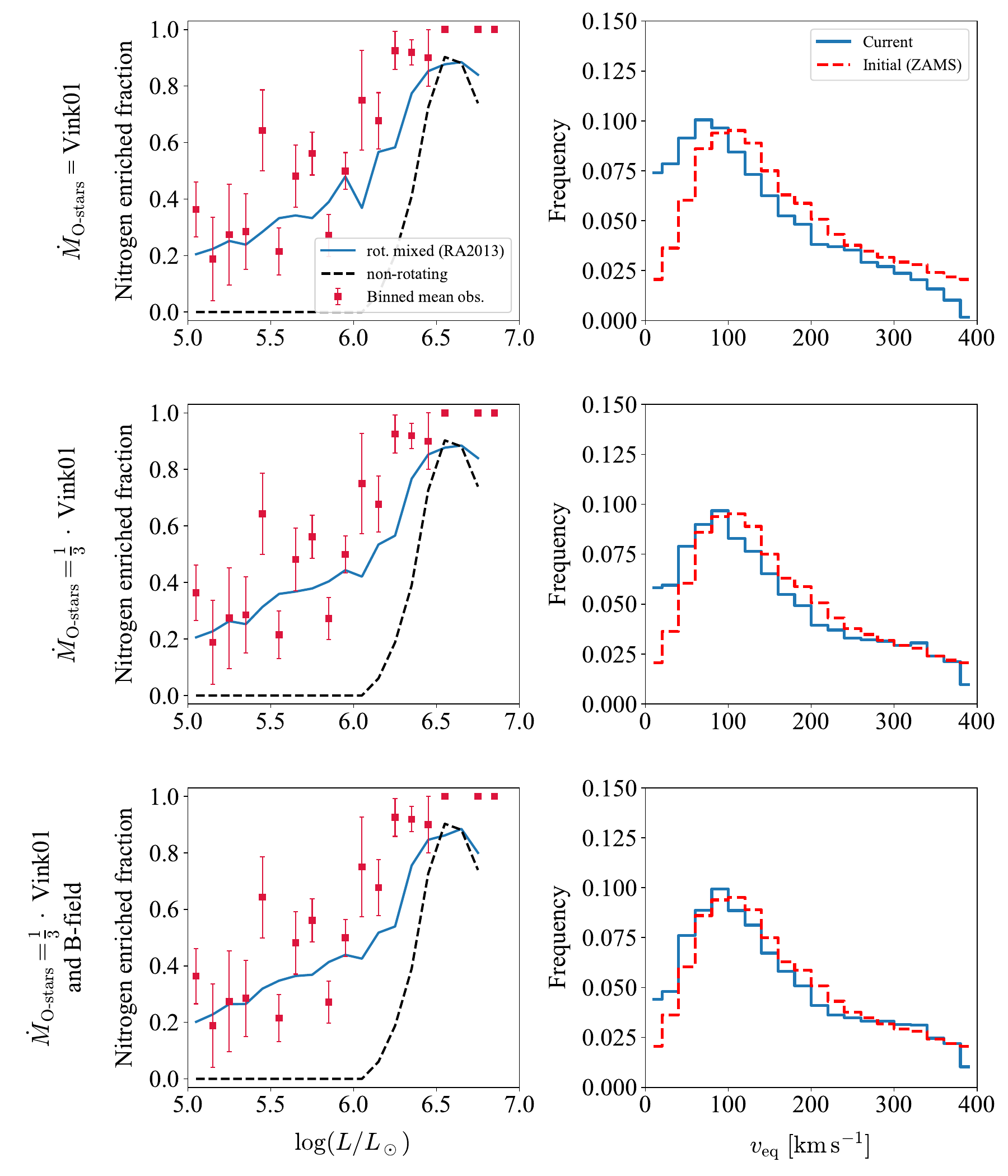}   \caption{Nitrogen-enriched probability as a function of luminosity compared with population synthesis predictions (\textit{left}), and the rotational velocity distribution of our population synthesis models (\textit{right}). The three rows correspond to different assumptions on the mass-loss rates and on whether the magnetic dynamo is switched on or off. Lines and symbols in the left sub-panels have the same meaning as in Fig.~\ref{fig: nitrogen_enriched_fraction}. The right sub-panels compare the evolved (solid blue) and initial (dashed red) velocity distributions, showing that the models do not spin up to velocities that are not observed.}  
\label{fig: compare_all}
\end{figure}

In Fig.~\ref{fig: compare_all} we compare the \emph{evolved} surface rotational velocities of our synthetic population against the~\citet{RA2013} input distribution. First, for the standard O-star rates of~\citet{Vink2001}, the stars spin down during evolution owing to both mass loss and the increase in radius during the MS, shifting the peak of the velocity distribution to lower velocities by roughly $40\,\mathrm{km\,s^{-1}}$.

Next, we reduce the~\citet{Vink2001} O-star rates by a factor of three. A lower mass-loss rate combined with a sufficiently high initial rotational velocity could in principle spin stars up during the MS. However, we find that the N-enriched probabilities remain largely unchanged, and although the loss of angular momentum is now lower, the models on average still do not spin up to unobserved velocities but instead typically spin down by roughly $20\,\mathrm{km\,s^{-1}}$.

Finally, for this reduced mass-loss scenario, we additionally test the impact of introducing an ST dynamo, which couples the core and the envelope and enables more efficient angular momentum transport between them. We again check whether the stars spin up to occupy the high-velocity tail, and find that the evolved velocity distribution remains largely unchanged compared to the initial one.

\section{MESA modelling details}
\label{app: mesa_details}

Here we provide additional details pertaining to the MESA evolutionary code. We use MESA version 12115 to construct our grid of massive and very massive stars. For the LMC-like composition models, the initial abundances of C, N, O, Mg, Si and Fe are taken from \citet{Brott2011}, while all other metals follow the solar mixture of \citet{Asplund2005} scaled down by 0.4\,dex.

Convection is treated using the standard mixing length theory (MLT) of~\citet{MLT68} with a fixed convective mixing length parameter $\alpha_\mathrm{MLT} = 1.5$. The MLT++ routine is switched off for the entire MS evolution, meaning the temperature gradient predicted by MLT is used as is, without any artificial reduction. Convective stability is determined using the Ledoux criterion, with diffusive semiconvective mixing following~\citet{Langer1983} and an efficiency of $\alpha_\mathrm{sc} = 1$. The different rotationally-induced instabilities and their diffusive implementation follow~\citet{Heger2000}.

For the reaction network we use the \texttt{basic.net} network~\citep{MESA11}, which consists of eight isotopes ($^{1}$H, $^{3}$He, $^{4}$He, $^{12}$C, $^{14}$N, $^{16}$O, $^{20}$Ne and $^{24}$Mg). To this we add $^{40}$Ca and $^{56}$Fe as inert species, giving ten isotopes in total. $^{40}$Ca is used as a fill-up species, absorbing the mass fraction of all metals not explicitly included in the network so that the total $Z$ is conserved, while the $^{56}$Fe abundance sets the metallicity scaling for the adopted wind mass-loss rates. The network is sufficient for following hydrogen burning through the pp chains and CNO cycle during the MS, the only phase considered here.

Opacities are taken from the OPAL Type 2 tables for the~\citet{Asplund2005} mixture, which account for changes in the mass fractions of the metals, mainly C and O. For the equation of state (EOS) we use the tables available in MESA, based mainly on the OPAL EOS tables~\citep{Rogers2002} together with a blend of other EOS tables~\citep[for further details, see][]{MESA11}.

\end{document}